\documentclass[reprint,aps,prd,notitlepage,superscriptaddress,nofootinbib]{revtex4-2}

\usepackage[utf8]{inputenc}
\usepackage[english]{babel}
\usepackage{amsmath}
\usepackage{amsfonts}
\usepackage{amssymb}
\usepackage{listings}
\usepackage{lipsum}
\usepackage{multirow}
\usepackage{datetime}
\usepackage{graphicx}
\usepackage{mathtools}
\usepackage{mathrsfs}
\usepackage{dcolumn}
\usepackage{multirow}
\usepackage[caption=false]{subfig}
\usepackage{soul}
\usepackage{breqn}
\usepackage{chngcntr}
\usepackage{color} 
\usepackage{hyperref}
\hypersetup{
    colorlinks=true, 
    pdfborder = {0 0 0.5 [3 3]},
    anchorcolor=black,
    citecolor=blue,
    linktoc=all,    
    linktocpage=true,
    linkcolor=red,
	urlcolor=blue
}

\usepackage{xcolor}

\begin{document}

\title{Nonlinear evolution of the spin-2 black hole superradiant instability
}

\author{Taillte May}
\email{tmay@perimeterinstitute.ca}
\affiliation{Perimeter Institute for Theoretical Physics\char`,{} Waterloo\char`,{} Ontario N2L 2Y5\char`,{} Canada}
\affiliation{Department of Physics and Astronomy\char`,{} University of Waterloo\char`,{} Waterloo\char`,{} Ontario N2L 3G1 Canada}

\author{William E. East}
\email{weast@perimeterinstitute.ca}
\affiliation{Perimeter Institute for Theoretical Physics\char`,{} Waterloo\char`,{} Ontario N2L 2Y5\char`,{} Canada}

\date{\today}

\begin{abstract} 
Massive spin-2 fields, which arise in proposed extensions of the standard model of particle physics and general relativity, give rise to a superradiant instability around spinning black holes. We study the nonlinear evolution of this instability in quadratic gravity, where quadratic curvature terms are added to the Einstein-Hilbert action. In this theory, we find that as the massive spin-2 field grows, the black hole's spin increases, accelerating the growth rate, and eventually leading to a superextremal horizon. This sharply contrasts with the spin-down and eventual saturation that occurs due to the backreaction of the superradiant instability of a minimally coupled massive spin-0 or spin-1 field. 

\end{abstract}

\maketitle

\section{Introduction}
General relativity (GR) predicts a massless spin-2 graviton as the mediator of gravity. However, massive spin-2 fields have been proposed 
in a number of contexts including modifications of GR \cite{Clifton:2011jh}, quantum theories of gravity,
and dark matter models~\cite{Aoki:2014cla,Babichev:2016bxi,Marzola:2017lbt,GonzalezAlbornoz:2017gbh,Jain:2021pnk,Manita:2022tkl,Kolb:2023dzp}. Among the theories giving rise to massive spin-2 field
are those that add quadratic curvature terms to the Einstein-Hilbert action (quadratic gravity)~\cite{Stelle:1976gc,Stelle:1977ry}, as well as theories that introduce an additional metric, like nonlinear massive gravity \cite{deRham:2010kj,deRham:2014zqa}, bimetric gravity \cite{Hassan:2011hr,Hassan:2011zd}, or other multimetric theories. 

Spinning black holes can trigger superradiant instabilities in the presence of massive bosonic fields ~\cite{1971JETPL..14..180Z,Misner:1972kx,Starobinsky:1973aij}. For spin-0 and spin-1 fields, the instability extracts energy and angular momentum from the black hole, spinning down the hole, and gradually reducing the instability rate, until a quasiequilibrium with a boson cloud is reached. 
The boson cloud slowly dissipates due to gravitational wave emission, while on even longer timescales higher mode clouds can grow, spinning down the black hole further.
This phenomenon has been used to constrain the existence of beyond the Standard Model physics through the observation of black hole spin measurements as well as searches for the gravitational waves sourced by the oscillating clouds \cite{Arvanitaki:2014wva,Arvanitaki:2016qwi,Cardoso:2018tly,Ng:2020ruv,Baryakhtar:2017ngi,Aswathi:2025nxa,Caputo:2025oap,Dergachev:2019oyu,Palomba:2019vxe,Collaviti:2024mvh,Tsukada:2018mbp,Tsukada:2020lgt,LIGOScientific:2025csr}. 

The nonlinear development of black hole superradiant instabilities has been studied in several scenarios.
A reflecting boundary, which arises naturally in the context of an asymptotitcally Anti--de Sitter spacetime, can 
lead to instability for a charged scalar field around a charged black hole~\cite{Bosch:2016vcp,Sanchis-Gual:2015lje} or for gravitational waves around a spinning black hole~\cite{Chesler:2018txn,Chesler:2021ehz}.
Studies in those settings found that their respective instabilities eventually saturate, resulting in a quasistable final state (though in some cases other instabilities arise on longer timescales). Similarly, the minimally coupled massive vector case was studied nonlinearly in Refs.~\cite{East:2017ovw,East:2018glu}, and the evolution and saturation of the instability was found to be quasiadiabatic to a good approximation, even for relativistic cases or when the cloud grows to order $10\%$ of the black hole mass. Given this holds for the vector case, we expect the quasiadiabatic approximation to be even better in the
minimally coupled, massive scalar case, since the associated superradiant timescales are even longer. 

This picture can change, however, when considering nonlinear effects besides gravity. Additional
bosonic interactions or coupling to other matter can lead to radiation, black hole absorption, or other sources of dissipation that halt the exponential 
growth of the cloud. This includes nonlinear interactions for a scalar boson~\cite{Gruzinov:2016hcq,Baryakhtar:2020gao,Omiya:2022gwu,Fukuda:2019ewf} or a coupling to charged matter in the vector case~\cite{Siemonsen:2022ivj,Caputo:2021efm,Fukuda:2019ewf}. 
In the interacting scalar case~\cite{Arvanitaki:2010sy,Yoshino:2015nsa}, or the Higgs-Abelian model of a massive vector~\cite{East:2022ppo,East:2022rsi,Brzeminski:2024drp}, nonlinear effects can become strong enough to lead to a explosive phenomenon referred to as a bosenova that disrupts the growth of the cloud (though in the scalar case, dissipative effects can hinder these values from being reached~\cite{Baryakhtar:2020gao,Omiya:2022gwu}).

The nonlinear regime of spin-2 superradiance has not been explored. 
In this case, there is no minimal way to couple such a field to gravity, and additional interactions (depending on the particular theory chosen) will necessarily be present. In addition, the linear spin-2 superradiant instability is parametrically faster than the scalar or vector case. For these reasons, one does not necessarily expect the phenomenology of the nonlinear evolution to be as simple as in the minimally coupled lower boson spin cases. Determining what happens in the spin-2 case is not only theoretically interesting in the context of understanding nonlinear superradiance, but an important step in attempting to use black hole observations to probe the existence of such fields.

Here, we investigate the behavior of massive spin-2 black hole superradiance using full nonlinear simulations of quadratic gravity. We choose this particular theory because it admits any vacuum GR solution as a solution, while also containing a massive spin-2 degree of freedom~\cite{Stelle:1977ry} and having a known well-posed initial value formation~\cite{Noakes:1983xd}. In contrast to the other cases described above, we find that nonlinear effects do not hinder the growth of the superradiant instability. 
The growing spin-2 cloud increases the black hole spin, accelerating the instability, and eventually leading to a superextremal apparent horizon. 

\section{Linearized spin-2 Superradiance}
The linear evolution of a massive spin-2 field $H_{ab}$ with mass $\mu$ on a Ricci-flat spacetime is described by the equations of motion \cite{Buchbinder:1999ar,Mazuet:2018ysa,Brito:2013wya},
\begin{align}\label{eq:lineareom}
    \nabla^c\nabla_c H_{ab} &= \mu^2H_{ab}-2R_{acbd}H^{cd}, &H^a_a = \nabla_aH^{ab}=0,
\end{align}
where $\nabla_a$ is the usual covariant derivative operator. 
Unless otherwise stated, we assume units with $G=c=1$ throughout.
As in the spin-0 and spin-1 cases, there is a linear superradiant instability in the spin-2 massive boson around a Kerr black hole with nonzero spin. 
For a mode solution (in, e.g., Boyer-Lindquist coordinates) of the form
\begin{align}
    H_{ab}(t, r, \theta, \varphi) = h_{ab}(r,\theta)e^{-i(\omega t - m\varphi)},
\end{align}
where $\omega = \omega_R + i \omega_I\in \Bbb{C}$ and $m\in \Bbb{Z}$, the superradiantly unstable modes satisfy the condition $0<\omega<m\Omega_H$ where $\Omega_H$ is the horizon angular velocity. 
The linear superradiant instability has been studied perturbatively in the $M_{\text{BH}}\mu\ll1$ limit \cite{Brito:2013wya}, and with $M_{\text{BH}}\mu\leq 0.8$ for the fastest growing superradiant mode \cite{Dias:2023ynv}, where $M_{\text{BH}}$ is the black hole mass and $\mu$ is the spin-2 boson mass. In Ref.~\cite{East:2023nsk}, the linear instability rates were calculated in the relativistic part of the parameter space for dimensionless spins $\chi\leq 0.998$ and $M_{\text{BH}}\mu\leq 2$.

In addition to the superradiant instability (and in contrast to the spin-0 and spin-1 cases), massive spin-2 fields exhibit an axisymmetric ($m=0$) linear instability around even nonspinning black holes when $\alpha:=\mu M$ (where $M$ is the spacetime mass) is sufficiently small. This instability was first found in Ref.~\cite{Babichev:2013una}, and has been studied in Refs.~\cite{Brito:2013wya,Myung:2013doa,Lu:2017kzi,Held:2022abx,East:2023nsk,Held:2021pht,Held:2025ckb}. In much of the space of boson masses and black hole parameters, this instability is faster than the superradiant instability. However, for sufficiently large $\alpha$, the $m=0$ instability becomes negligible or absent and superradiant instabilities dominate. 
For example, when $\chi=0.99$, the $m=0$ instability
dominates for $\alpha\lesssim1.4$~\cite{East:2023nsk}.

\section{Nonlinear Theory}
While the equations of motion for a massive spin-2 field on a black hole background are unique at the linear level, there is no unique, minimal way to nonlinearly couple such a field to gravity. One nonlinear theory which leads to a well-posed initial value problem is quadratic gravity, which has the following action:
\begin{align}
\label{eq:quad_acion}
    S = \int d^4x\sqrt{-g}\left(R-\frac{1}{2\mu^2}C^{abcd}C_{abcd}+\frac{1}{6 m_0^2}R^2\right) \ ,
\end{align}
where $g$ is metric determinant, $R$ is the Ricci scalar, and $C_{abcd}$ is the Weyl tensor.
This theory, also known as Stelle gravity, admits any vacuum solution to the Einstein equations as a solution, and is renormalizable around flat space~\cite{Stelle:1976gc}. However, as a fourth-order (nondegenerate) theory, Ostrogadsky's theorem implies the Hamiltonian is unbounded from below~\cite{Woodard:2015zca} with the massive spin-2 degree of freedom acting as a ghost.
Of particular relevance for this study is that this theory can still be formulated in a way that leads to well-posed initial value problem~\cite{Noakes:1983xd,Morales:2018imi}, which means we can use it to address questions regarding the backreaction of the superradiant instability.
An alternative nonlinear theory with a massive spin-2 field is massive bigravity~\cite{Hassan:2011hr,Hassan:2011zd}, though in that case, general well-posed evolution schemes are currently lacking~\cite{deRham:2023ngf,Kozuszek:2024vyb}.  

In quadratic gravity, the traceless part of the Ricci tensor $\tilde{R}_{ab}$ acts
as a spin-2 degree of freedom with mass $\mu$, while $R$ obeys the scalar wave equation with mass $m_0$. In particular, linearizing perturbations on a Ricci-flat solution gives Eq.~\eqref{eq:lineareom}, with $H_{ab}=\tilde{R}_{ab}$ \cite{Myung:2013doa}. Since in this study we are beginning with a Ricci-flat background solution, and the scalar superradiant instability operates on much longer timescales than we consider here, we expect that $R$ should remain negligible regardless of its mass. Therefore, we assume $R=0$ below (or equivalently, consider Einstein-Weyl gravity).

As mentioned above, in addition to the superradiant instability, spin-2 bosons exhibit an axisymmetric ($m=0$) instability.
In Ref.~\cite{East:2023nsk}, it was shown that, depending on the black hole parameters and the sign of the initial perturbation, the nonlinear evolution of this instability in Einstein-Weyl gravity results in either a superspinning black hole, a zero mass naked singularity formed in finite time, or simply another black hole with a larger mass and spin surrounded by a cloud of the spin-2 field.
Here, we show that even in the region of the parameter space where there is no $m=0$ instability, 
superradiant instabilities can also result in 
superspinning apparent horizons.

\section{Methods}
We investigate the nonlinear evolution and backreaction of the spin-2 superradiant
instability by numerically evolving quadratic gravity as in Ref.~\cite{East:2023nsk}. See Appendix~\ref{sec:num_scheme} for more details.
We begin from initial conditions describing a Kerr black hole perturbed by a superradiantly unstable mode. We fix the initial dimensionless spin of the black hole to be $\chi=0.99$, and consider two values for the spin-2 mass: $\mu M= 1.5$ and $1.9$, where $M$ is the initial black hole mass. For these parameters, the fastest growing superradiant modes are the $m=2$ and $m=3$, with instability rates $\omega_I = 1.6\times 10^{-3}/M$ and $1.2\times 10^{-4}/M$, respectively (the $m=0$ and 1 modes are stable).

We construct the desired superradiant mode solutions by evolving a massive spin-2 perturbation on the black hole background for several e-foldings. After this time, the test-field profile is strongly dominated by the fastest growing superradiant instability. The test-field data is then rescaled as desired, and is used as initial data for full nonlinear evolution. We check that the errors introduced by starting with such test-field initial data is sufficiently small by considering several different amplitudes.
For the evolution simulations, we use a grid spacing of $dx=0.02M$ around the black hole, though we have
also considered a lower resolution to check numerical convergence. Assuming third order convergence, we estimate an error in the instability rate of $\approx 15\%$, and an error in the black hole physical parameters $M_{\rm BH}$ and $J_{\rm BH}/A_{\rm BH}$ of $\approx4\%$.
More details regarding these are given in Appendices \ref{sec:amplitude_dependence} and \ref{sec:resolution_study} .

Since we are restricting to solutions with $R=0$, $\tilde{R}_{ab}$ is equal to the Einstein tensor, and is conserved $\nabla_a\tilde{R}^{ab}=0$ through the Bianchi identity.
It is therefore instructive to treat it analogously to the stress energy tensor
and define an effective energy and angular momentum, decomposed into
azimuthal components, as
\begin{align}\label{eq:EJm_def}
    E^{m=m'} = \frac{1}{8\pi}\int d^3x\sqrt{\gamma} n^bt^a\tilde{R}_{ab}e^{-im'\varphi},\nonumber\\
    J^{m=m'} = \frac{1}{8\pi}\int d^3x\sqrt{\gamma} n^b\varphi^a\tilde{R}_{ab}e^{-im'\varphi}.
\end{align}
Here $E:=E^{m=0}$ and  $J:=J^{m=0}$ correspond to the total energy and angular momentum, respectively, and higher $m$ components encode the azimuthal dependence of the effective stress energy tensor.
In Eq.~\eqref{eq:EJm_def}, $\gamma$ is the determinant of the spatial metric, $n^a$ is the normal to the spacelike hypersurface, $t^a$ is the Killing vector associated with the stationary background spacetime, and $\varphi^a$ is the Killing vector associated with the background axisymmetry. At the linear level, $E$ and $J$ are conserved up to fluxes through the black hole horizon. Furthermore, for superradiant mode solutions (where $\tilde{R}_{ab}\sim e^{im\varphi}$ with $m\neq0$), these quantities are both zero at the linear level.
This means that a study of the nonlinear behavior of the initial perturbation is needed to determine the leading order backreaction on the black hole's mass and angular momentum. 

In order to better understand the behavior of the solutions,  $E^{m}$ and $J^{m}$ with $m\neq 0$ are calculated.
Those modes can be used to determine which superradiant modes are growing, as well as the complex frequencies of those modes. 

During the evolution, we also track the black hole apparent horizon and calculate its area $A_{\rm BH}$ and associated angular momentum $J_{\rm BH}$, using this to define a mass $M_{\rm BH}$ through the Christodoulou formula (see Appendix \ref{sec:num_scheme} for details). We note that we calculate the horizon angular momentum with respect to the coordinate direction of the original black hole spin, and there can be gauge dependence to this quantity when the spacetime differs significantly from axisymmetry.

\section{Results}
We numerically evolve the full quadratic gravity equations for several superradiant spin-2 perturbations to a black hole with initial dimensionless spin $\chi=0.99$: $m=2$ with $\alpha=1.5$ and $m=3$ with $\alpha=1.9$. We find that as the massive spin-2 cloud grows, the black hole spin and the cloud growth rate both increase. For $m=2$ with $\alpha=1.5$, where (due to the faster instability rate) we follow the growth of the superradiant instability as it strongly backreacts, we find that the black hole apparent horizon eventually becomes superextremal.

We begin by describing the properties of the superradiant perturbations during the weakly nonlinear growth phase.
To linear order in the perturbation amplitude, the effective energy and angular momentum defined in Eq.~\eqref{eq:EJm_def} is zero, but from the evolution we can extract the leading order (quadratic in the amplitude) contribution to these quantities. We find that both $E$ and $J$ are negative in all the cases we consider.

\begin{figure}[h]
    \centering
    \includegraphics[width=1.0\linewidth]{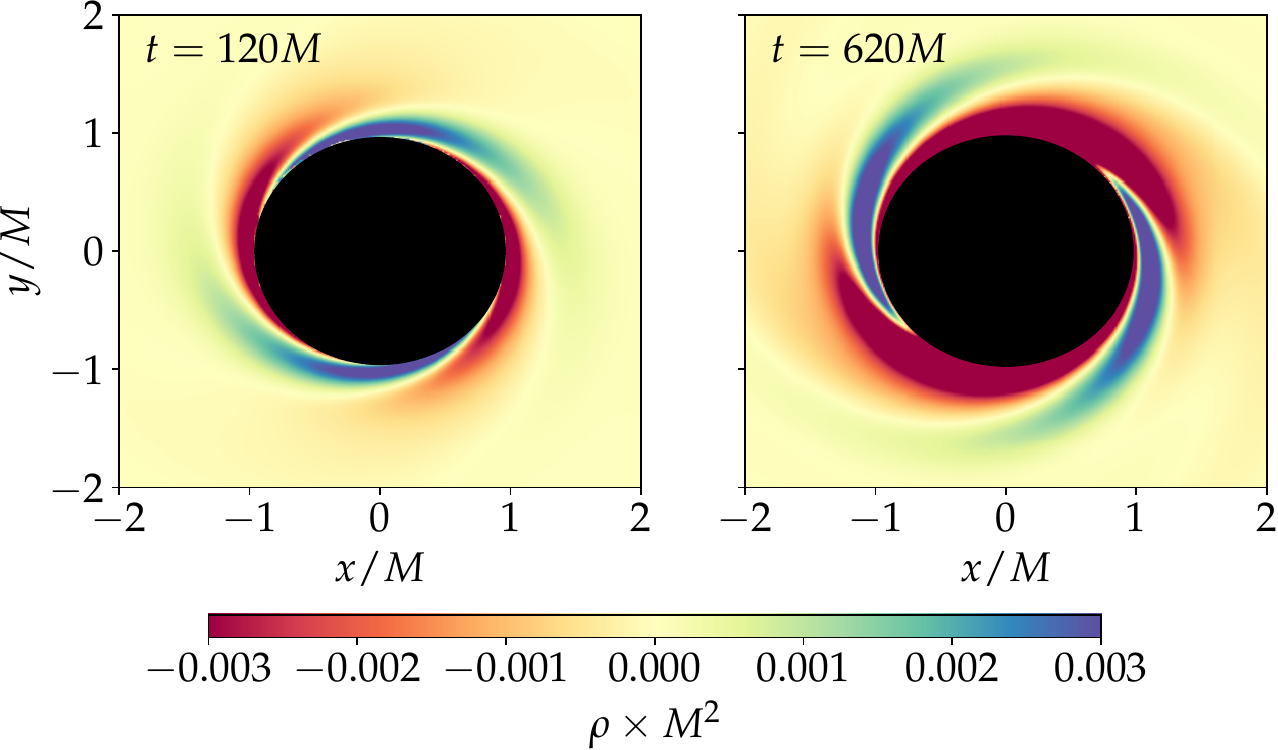}  
  \caption{The profile of the effective energy density $\rho=n^bt^a\tilde{R}_{ab}/(8\pi)$ of an $m=2$, $\alpha=1.5$ perturbation on the equatorial ($z=0$) plane, where the black hole is at the origin, and the spin points out of the page. It can be seen that there are two negative and two positive energy peaks, consistent with the leading order $m=2$ azimuthal dependence.
  The left panel shows the field profile at early times, where the backreaction on the black hole is small. The right panel shows the field approximately one e-folding time later, where the overall negative monopolar component to the energy density is clearly visible.
  }
  \label{fig:snapshot_al1p5}
\end{figure}

This is illustrated in Fig. \ref{fig:snapshot_al1p5}, where we show an effective energy density $\rho=n^bt^a\tilde{R}_{ab}/(8\pi)$ for the initial perturbation with $m=2$ and $\alpha=1.5$ at two different times, approximately one e-folding time apart. At the earlier time, the nonlinear effects are small and the linear $m=2$ behavior is most apparent in the effective energy density. At the later time, where nonlinear effects are stronger, there is a noticeable skew toward negative values in the azimuthal angle dependence. 

For a wave with fixed frequency and azimuthal number, we expect the energy and angular momentum to obey $E/J=\omega_R/m$ (assuming any imaginary frequency component is small enough compared to the real component to be ignored). We find that this approximately holds for the superradiant spin-2 modes we study here and the quantities defined in Eq.~\eqref{eq:EJm_def}. In particular, we find that the value of $J/E$ oscillates around a constant value, $J/E\approx 2.3M$ and $2.5M$ for $(m,\alpha)=(2,1.5)$ and $(m,\alpha)=(3,1.9)$, respectively. This agrees to within $\approx 5\%$ with the value $m/\omega_R\approx2.4$ (for both cases) found from  either the linear calculation or from the frequency of oscillation of $E^{m=2}$.
Note, however, that for $(m,\alpha)=(2,1.5)$, the measured value of $J/E$ is consistent with being on either side of the threshold $1/\Omega_H\approx 2.31 M$ (for $\chi=0.99$), 
below (above) which, an adiabatic change to a Kerr black hole will change the mass with the same (opposite) sign as it changes the horizon area.

We have also looked for gravitational radiation sourced nonlinearly during the growth of the spin-2 field, but find that it is too small as to be reliably extracted from the simulation. Applying the quadrupole formula to the $\alpha=1.5$, $m=2$ case to obtain an order-of-magnitude estimate (and assuming a harmonic time dependence) 
indicates that any gravitational radiation is likely subdominant to the exponential tail of the spin-2 cloud even at a distance of $\sim100M$.

\begin{figure}[h]
\centering
\includegraphics[width=8cm]{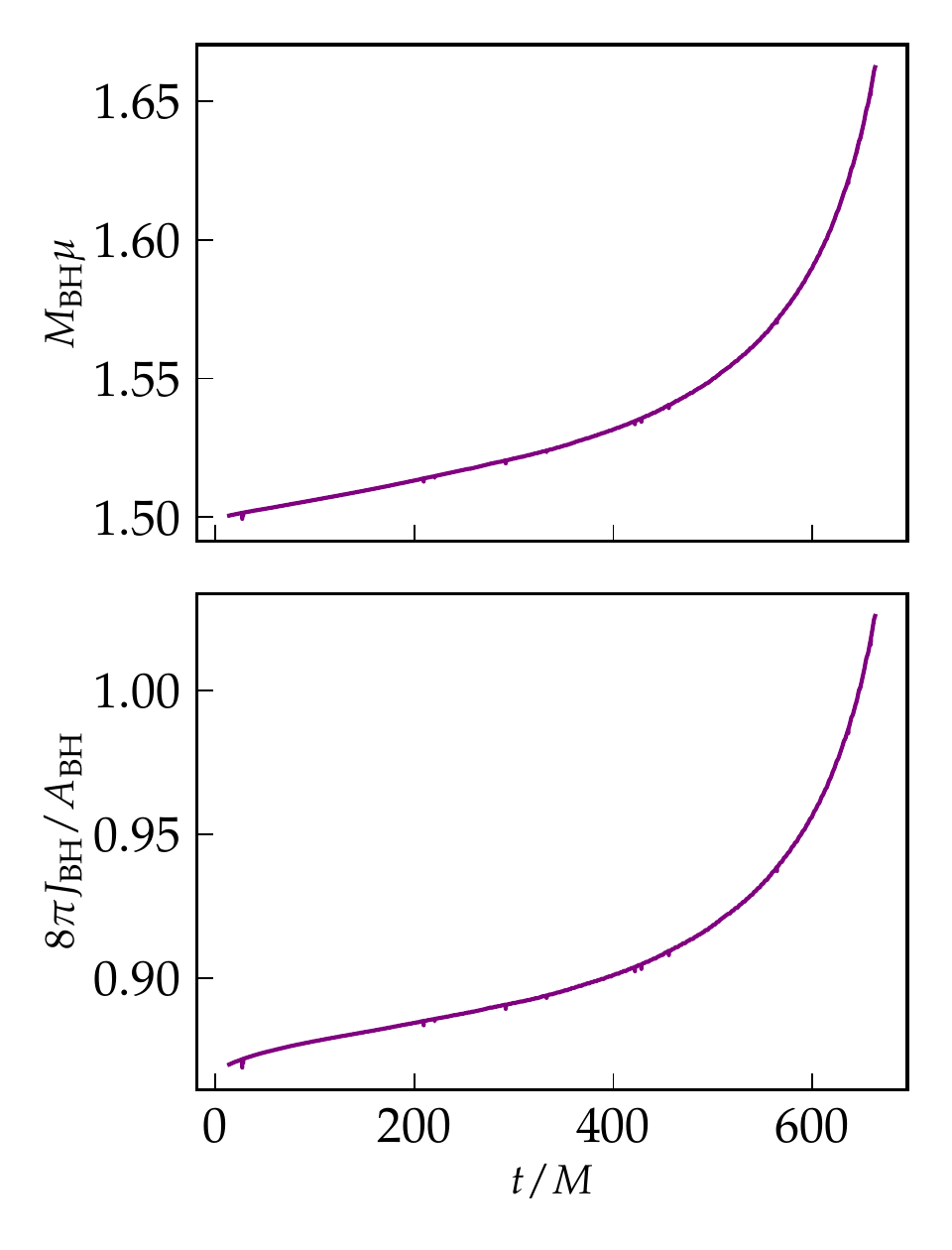}
\caption{The black hole apparent horizon mass (top) and angular momentum relative to its area (bottom) as a function of time. The spin increases with time, and shortly after it reaches extremality ($8\pi J=A_H$) we are no longer able to find an apparent horizon. 
}
\label{fig:mass_spin_al1p5}
\end{figure}

Focusing now on the $m=2$ and $\alpha=1.5$ case, we begin with a perturbation where $E^{m=2}\approx 6\times10^{-3}M$ and follow the superradiant growth as it transitions from the linear instability rate to strongly backreacting on the black hole. 
We track the mass and angular momentum of the black hole apparent horizon with time and find, consistent with the fact that $E$ and $J$ are negative, that both increase until an apparent horizon can no longer be found. As can be seen in Fig.~\ref{fig:mass_spin_al1p5}, shortly before this occurs, the apparent horizon becomes superextremal in the sense that the ratio of angular momentum to area satisfies $8\pi J_{\rm BH}/A_{\rm BH}>1$.

Though we cannot rule that it is an artifact of our horizon-finding algorithm, our failure to find an apparent horizon at late times, coupled with the rapidly growing spin of the horizon leading up to this, suggests that the apparent horizon may have disappeared. Without a trapped surface, one can no longer continue the evolution while excising the black hole interior. However, in the lead up to this, we see no sign of diverging curvature or other singularities outside the black hole horizon.

\begin{figure}[h]
\centering
\includegraphics[width=8cm]{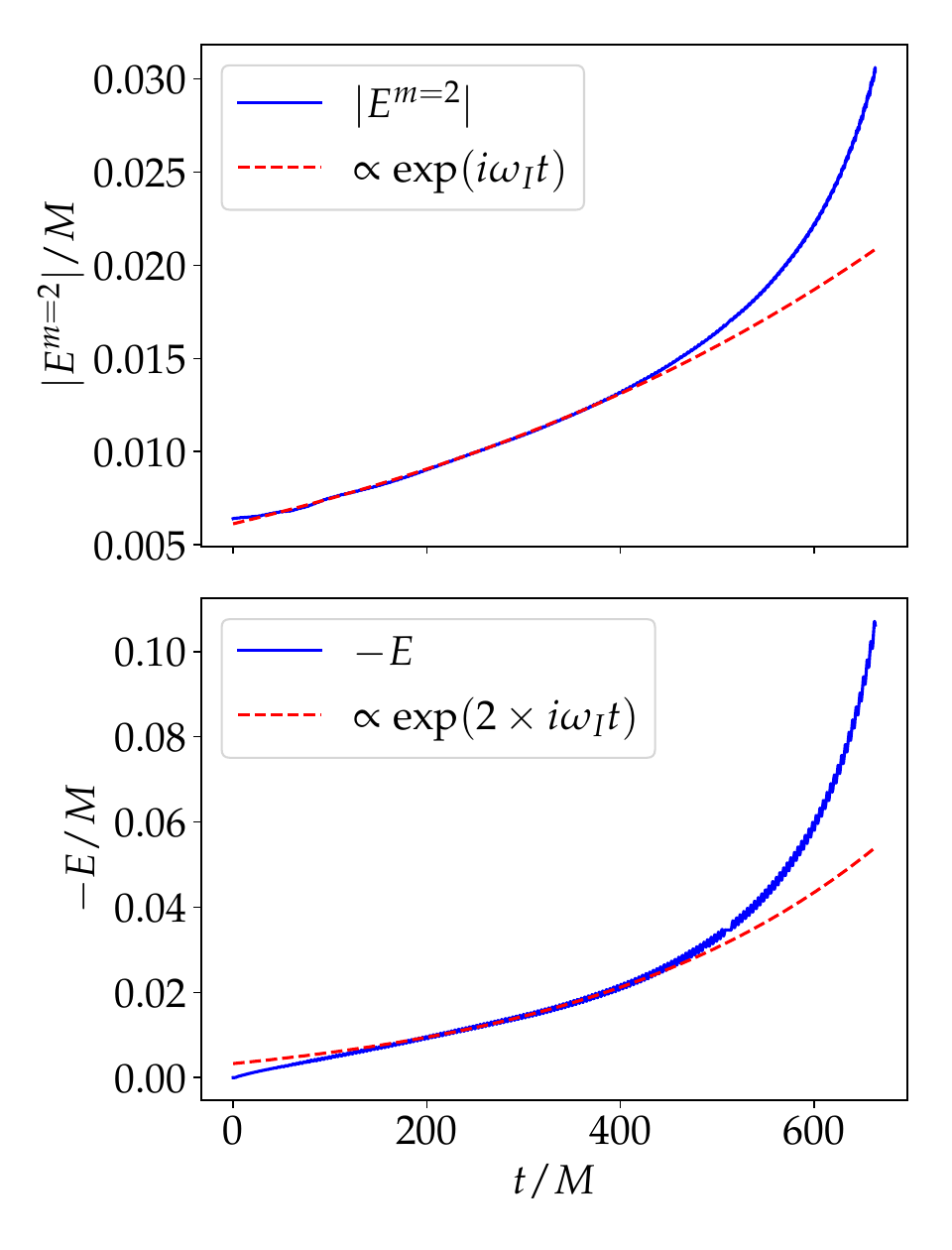}
\caption{The magnitude of the $m=2$ (top panel) and $m=0$ (i.e., total; bottom panel) azimuthal component of the effective energy density as a function of time for an initial perturbation with $m=2$ and $\alpha=1.5$. 
The linear instability growth rate for a background with $M_{\rm BH} \mu=1.5$ and $\chi=0.99$
is shown for comparison (dashed lines).
After a brief transient period due to the linear initial conditions, 
the total energy growth at intermediate times is consistent with a quadratic effect associated with the $m=2$ linear superradiant growth. 
At $t/M\approx500$, when the growth rate begins to noticeably differ from the linear prediction, the black hole has changed such that $M_{\rm BH} \mu=1.55$ and $\chi=0.9964$. 
}
\label{fig:growthE2al1.5}
\end{figure}

The growth of the black hole's mass and angular momentum closely
tracks that of the massive spin-2 cloud.
Examining the azimuthal modes of the effective energy density through the calculation of $E^{m=m'}$ according to Eq.~\eqref{eq:EJm_def},
we find that $m=2$ initially dominates over the other azimuthal modes, as expected from the linear instability when nonlinear effects are subdominant. 
The only other significant $m$ component in the effective energy density is $m=0$, which determines the total effective energy of the field. We show the growth with time of both of these quantities in Fig. \ref{fig:growthE2al1.5}. The rate at which the field's energy $E$ decreases at early times is
twice the superradiant growth rate, consistent with it having a quadratic dependence on the superradiantly unstable mode. 
Over the course of the evolution, the apparent horizon area also shows a net increase, corresponding to the ratio of the increase in the black hole's angular momentum to mass $\Delta J_{\rm BH}/\Delta M_{\rm BH}$ being slightly below $1/\Omega_H$.

From Fig.~\ref{fig:growthE2al1.5}, we can see that as the spin-2 cloud grows into the nonlinear regime, and as $M_{\rm BH}$ and $J_{\rm BH}$ increase, the growth rate of $E^{m=2}$ and $E$ increase significantly compared to the expected linear
instability rate. There could be several reasons for this.
If we treat the evolution quasiadiabatically, and imagine that we can still apply a linear instability analysis around a Kerr background to the evolving black hole, then as the black hole spin and mass increase,
this indicates that not only will the $m=2$ instability rate increase, 
but the $m=0$ instability will as well.
At $t/M\approx500$, the black hole mass and spin have changed such that $M_{\rm BH}=1.55/\mu$ and $\chi_{\rm BH}=0.9964$. The linear instability rates\footnote{We follow the same procedure as Ref.~\cite{East:2023nsk} to calculate linear instability rates.} for these values are $\omega_I M \approx 0.0051 $ and $0.0018$, for $m=0$ and 2, respectively. This means that the $m=0$ linear instability growth rate exceeds the quadratic growth in $E$ associated with the $m=2$ superradiant instability. Toward the end of the evolution, the growth in $E$ is roughly consistent with the increasing linear $m=0$ instability rate, though the growth in $E^{m=2}$ is much larger and not consistent with linear behavior. Of course, as the black hole begins to evolve rapidly, the linear, quasiadiabatic analysis need not apply, and certainly breaks down as the black hole becomes superextremal, so additional nonlinear effects may also be contributing to the enhanced growth rate at late times. 

Due to its significantly smaller instability rate, we only evolve the initial perturbation with $m=3$ and $\alpha=1.9$ for long enough to measure the values of $E$ and $J$ associated with the massive spin-2 cloud.
Nevertheless, the negative values of these quantities suggest that the nonlinear development of the instability
will also result in a superspinning black hole in finite time.
Furthermore, we note that in this case, based on the linear analysis around a Kerr black hole, we do not expect the $m=0$ instability to dominate over the $m=3$ instability, even as the black hole approaches extremal spin.

\section{Discussion and conclusion}
For a minimally coupled massive scalar or vector field, a superradiantly growing cloud decreases the mass and angular momentum of a spinning black hole in such a proportion that the horizon frequency decreases until the superradiant instability shuts off. 
For a massive spin-2 field, on the other hand, there is no canonical way to couple it to gravity, and no energy conditions to guarantee that this happens.
When self-interactions or couplings to other fields are sufficiently strong in the scalar or vector case, they can halt the exponential growth of the cloud before the instability of backreacts significantly on the black hole. In the massive tensor case, such interactions will be required, with their exact form depending on the particular nonlinear theory.  

In this work, using nonlinear evolutions in quadratic gravity, we find that in that theory the spin-2 superradiant instability does indeed grow until it backreacts on the black hole. However, in this case, it
spins up the black hole, leading to a run away effect whereby the horizon eventually becomes superextremal. The final object is not resolved numerically but is certainly not a Kerr black hole. 
This behavior resembles that found for the axisymmetric ($m=0$) massive spin-2 instability
when evolved nonlinearly in quadratic gravity for some values of black hole spin and initial perturbations~\cite{East:2023nsk}.

For future work, it would be interesting to determine whether the behavior found here occurs
in other nonlinear theories with massive spin-2 fields, for example, if bimetric gravity~\cite{Hassan:2011zd} could be formulated in a way where the nonlinear development of this instability could be addressed.
While nominally ghost-free, in contrast to quadratic gravity, this class of theories also generically violates the equivalent of the null energy condition~\cite{Baccetti:2012re}. Furthermore, for bimetric gravity in spherical symmetry, the backreaction of the $m=0$ massive spin-2 instability can either decrease or increase the area of the black hole horizon, depending on the sign of the perturbation~\cite{Zhong_in_prep}, similar to quadratic gravity.

Previous works have considered how one can constrain massive spin-2 fields through the superradiant instability using spin measurements or gravitational wave searches, assuming the instability saturates by spinning down the black hole in a manner similar to the massive spin-0 or spin-1 cases~\cite{Brito:2013wya,Brito:2020lup,Dias:2023ynv} (with perturbative analyses restricted to the regime where the superradiant instability is subdominant to the $m=0$ instability).
In the cases studied here in quadratic gravity, the backreaction of superradiance is quite different; in particular, we do not find that the instability saturates with a Kerr-like black hole.
Though this makes it harder to judge the expected phenomenological signatures, it does suggest that observations of Kerr-like black holes can be used to rule out such theories where the Compton wavelength of the massive spin-2 field is commensurate with the black hole radius, as these instability timescale is quite short. Restoring astrophysical units, we recall that the cases studied here correspond to $\mu\sim 2\times 10^{-11}\text{ eV} (10\ M_{\odot}/M_{\rm BH})$ with instability e-folding times of $0.03$--$0.41$ sec $M_{\rm BH}/(10\ M_{\odot})$. 

We approximate the maximum value of $\alpha$ for which a black hole with spin $a/M = 0.99$ and mass $M=O(M_\odot)$ is unstable on the timescale of the age of the universe. We use the results of \cite{East:2023nsk}, where we fit the instability timescale for the $m=1,2,3$ modes to be $\tau(m) = \exp(-8.9 + 2.1 m)$ seconds (specifically for $M_{\text{BH}} = 2M_\odot$), and extrapolate to higher $m$ modes. Then we use \cite{Brito:2013wya} for the values of $\alpha$ for maximally efficient superradiant scattering, to find that these black holes are unstable, at a timescale $\tau$ equal to the age of the universe, for $\alpha \lesssim 7.5$, or for $\mu \lesssim  O(10^{-9})$eV. 
The upper mass scale estimated here relies on extrapolating the available low-\(m\) instability rates and should not be interpreted as a demonstrated stability boundary for all astrophysical Kerr black holes. This result does not establish constraints on the much heavier spin-2 degrees of freedom considered in high-energy applications of quadratic gravity. An observational exclusion window would also require an understanding of the nonlinear endpoint of the instability.

On the low $\alpha$ or low boson mass side, the $m=0$ monopole instability will always dominate and its instability timescale is $\tau \approx \sqrt{2} M_{\text{BH}}/\alpha$ \cite{Camps:2010br}. Using this approximation, the smallest boson masses for which a black hole of mass $M_{\text{BH}} = 100M_\odot$ would be unstable on the timescale of the age of the universe are $\mu \gtrsim 10^{-33}$eV.

\acknowledgments
Research at Perimeter Institute is supported in part by
the Government of Canada through the Department of
Innovation, Science and Economic Development Canada
and by the Province of Ontario through the Ministry of
Colleges and Universities. The authors acknowledge support
from a Natural Sciences and Engineering Research Council of Canada Discovery Grant and an Ontario Ministry
of Colleges and Universities Early Researcher Award.
This research was enabled in part by support provided
by the Digital Research
Alliance of Canada (alliancecan.ca). Calculations were
performed on the Symmetry cluster at Perimeter Institute and the Narwhal cluster at the École de technologie supérieure in Montreal.

\section*{DATA AVAILABILITY}
The data that support the findings of this article are not publicly available upon publication because it is not technically feasible and/or the cost of preparing, depositing, and hosting the data would be prohibitive within the terms of this research project. The data are available from the authors upon reasonable request.

\bibliography{bib.bib}

\appendix

\section{NUMERICAL SCHEME}
\label{sec:num_scheme}

We numerically evolve the Einstein-Weyl equations, or equivalently, the quadratic gravity 
equations with $R\equiv 0$, using 
$\{g_{ab},\partial_tg_{ab},R_{ab},\partial_tR_{ab}\}$ as the evolution variables. We use the general harmonic formulation \cite{Noakes:1983xd}, where the coordinates $x^a$ satisfy $\Box x^a=H^a$ for specified functions $H^a(g_{ab})$. (See Ref.~\cite{Held:2023aap} for an alternative formulation.) The metric evolution equation is
\begin{align}
    \label{eq:metric_evo}
    g^{cd}\partial_c\partial_dg_{ab}&= -2\partial_{(a}H_{b)}+2H_c\Gamma^c_{ab}-2\Gamma^c_{da}\Gamma^d_{cb} \\
    &-\kappa\left(n_aC_b+n_bC_a-n_cC^cg_{ab}\right)\\
    &-2\partial_cg_{d(a}\partial_{b)}g^{cd} -2R_{ab}.
\end{align}
Here $\Gamma_{ab}^c$ are the usual Christoffel symbols, and $C^a=H^a-\Box x^a$ is the harmonic constraint \cite{Gundlach:2005eh}, with $\kappa\neq 0$ corresponding to the inclusion of constraint damping terms.

The Ricci tensor evolution equation is
\begin{align}
    \label{eq:Rabevo_EW}
    \Box \tilde{R}_{ab} &= \mu^2\tilde{R}_{ab} - 2R_{acbd}\tilde{R}^{cd}+\frac{1}{2}g_{ab}\tilde{R}^{cd}\tilde{R}_{cd} \\
    &-\kappa \left( n_a \nabla^c \tilde{R}_{bc} + n_b \nabla^c \tilde{R}_{ac} - \frac{1}{2} g_{ab} n^c \nabla^d \tilde{R}_{cd} \right),
\end{align}
where $R_{abcd}$ is the Riemann tensor, calculated from $g_{ab},\partial_c g_{ab}$, and $\partial_c\partial_dg_{ab}$, where $\partial_t^2g_{ab}(g_{ab},\partial_tg_{ab})$ is calculated according to Eq. \eqref{eq:metric_evo}. We have added a term proportional to $\kappa$ to damp violations of $\nabla_a\tilde{R}^{ab}=0$ on a timescale of $1/\kappa$. Here we typically set $\kappa \sim 1/M_{\text{BH}}$.

We begin the nonlinear evolution using a black hole in Cartesian Kerr-Schild coordinates, and use the gauge $H^a=-\left(g^{bc}\Gamma^a_{bc}\right)_{\rm Kerr}$, where the gauge functions $H^a$ are stationary, and are fixed by the choice of the Kerr background. 
Spatial derivatives are calculated using a fourth-order finite difference scheme, and the time evolution uses the fourth-order Runge Kutta algorithm. We use mesh refinement, meaning that there is higher numerical resolution in the vicinity of the black hole. However, the interpolation of the refinement boundaries only converges at third-order with temporal resolution.

The evolution code calculates the position of the black hole's apparent horizon using a flow method \cite{Pretorius:2004jg}, finding the outermost marginally trapped surface. Once the black hole apparent horizon is found, the associated angular momentum is calculated from the contraction of the extrinsic curvature $K_{ij}$ with the approximate rotational Killing vector $\varphi_i$ \cite{Ashtekar:2004cn},
\begin{align}
    J_{\rm BH} = \frac{1}{8\pi}\int \varphi_iK^{ij}dA_j.
\end{align}
The Christodoulou mass associated with the horizon is then calculated using the area of the apparent horizon $A_{\rm BH}$, and this angular momentum,
\begin{align}
    M_{\rm BH} = \sqrt{\frac{A_{\rm BH}}{16\pi}+\frac{4\pi J_{\rm BH}^2}{A_{\rm BH}}}.
\end{align}
As noted in the main text, since in this case $\varphi_i$ is only an approximate rotational Killing vector, $J_{\rm BH}$ as defined here may be mildly coordinate dependent.
Following Ref.~\cite{Pretorius:2004jg}, we excise a region inside the apparent horizon from the numerical domain. This region is dynamically adjusted as the horizon evolves.

We use Cartesian compactified coordinates, with seven levels of mesh refinement, each having approximately $251^3$ points for the results shown in the main text. The finest level covers the region $x,y,z\in[-2.5M,2.5M]$, 
where the black hole is centered at the origin and initially has polar radius $\approx1.14M$ and equatorial radius $\approx 1.51M$. 

\section{DEPENDENCE ON INITIAL PERTURBATION AMPLITUDE}
\label{sec:amplitude_dependence}

We construct initial data for the nonlinear evolution of the full Einstein-Weyl equations corresponding to the fastest growing superradiant mode. We do this by evolving the linearized equations~\eqref{eq:lineareom} with a seed perturbation until the fastest growing superradiant mode dominants and all other modes are negligible. This solution to the linearized equations can then be scaled and used as initial data for the full nonlinear evolution. 
The initial perturbation for the nonlinear evolution should ideally be large enough such that the entire evolution can be simulated in a reasonable amount of computer time, while also being small enough that its initial backreaction has a sufficiently small effect on the dynamics and the late time evolution is insensitive to the initial amplitude.

\begin{figure}[h]
\centering
\includegraphics[width=8cm]{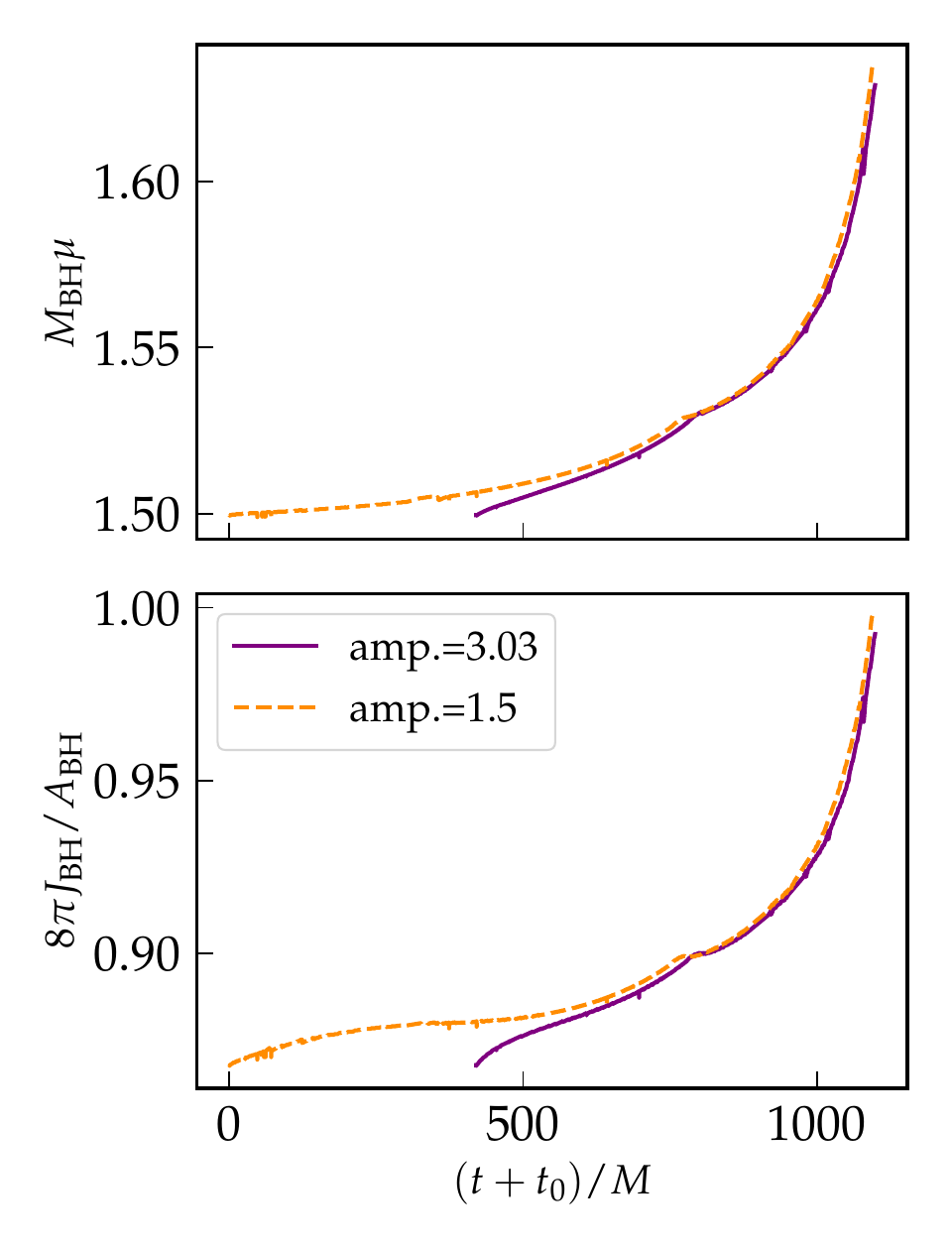}
\caption{The apparent horizon parameters are plotted for two cases, where the initial perturbation (with $m=2$ and $\alpha=1.5$) is scaled according to the factor ``amp.'' before beginning the full nonlinear evolution. A time shift $t_0=420M$ is introduced in the larger amplitude case to better demonstrate the similarity in evolution. While there are some effects at early times due to the change from linear to nonlinear evolution, the phenomenology is the same and the effect on the late time behavior is not significant. There is a visible kink in the evolution at $t/M\approx800$. For the simulations in this plot the excision surface was not changed until this time, where excision surface is too close to the apparent horizon for a short window $750\lesssim t/M \lesssim 800$, affecting the apparent horizon solver. This problem is avoided in the simulation shown in the main text by changing the excision surface at earlier times so that the distance between the excision boundary and the apparent horizon is always sufficient.
}
\label{fig:diff_amps}
\end{figure}

We compare evolutions with different initial amplitudes to verify that the initial backreaction does not meaningfully change the phenomenology. In Fig. \ref{fig:diff_amps}, we show the evolution of two different initial amplitudes, differing by a factor of two ($2.02$ to be exact), where the larger value corresponds to the amplitude used in the main text. From the figure, we see that this increase in initial amplitude, or delay in time before beginning the nonlinear evolution, does not strongly impact the late time behavior. The black hole still spins up until the apparent horizon is no longer found. 

\section{RESOLUTION STUDY}
\label{sec:resolution_study}

We use a fourth order evolution code described in Ref.~\cite{East:2023nsk} (but without restricting to axisymmetry) to solve the dynamics. We compare evolution with different resolutions to check convergence and evaluate the numerical error in the simulation. 

\begin{figure}[h]
\centering
\includegraphics[width=8cm]{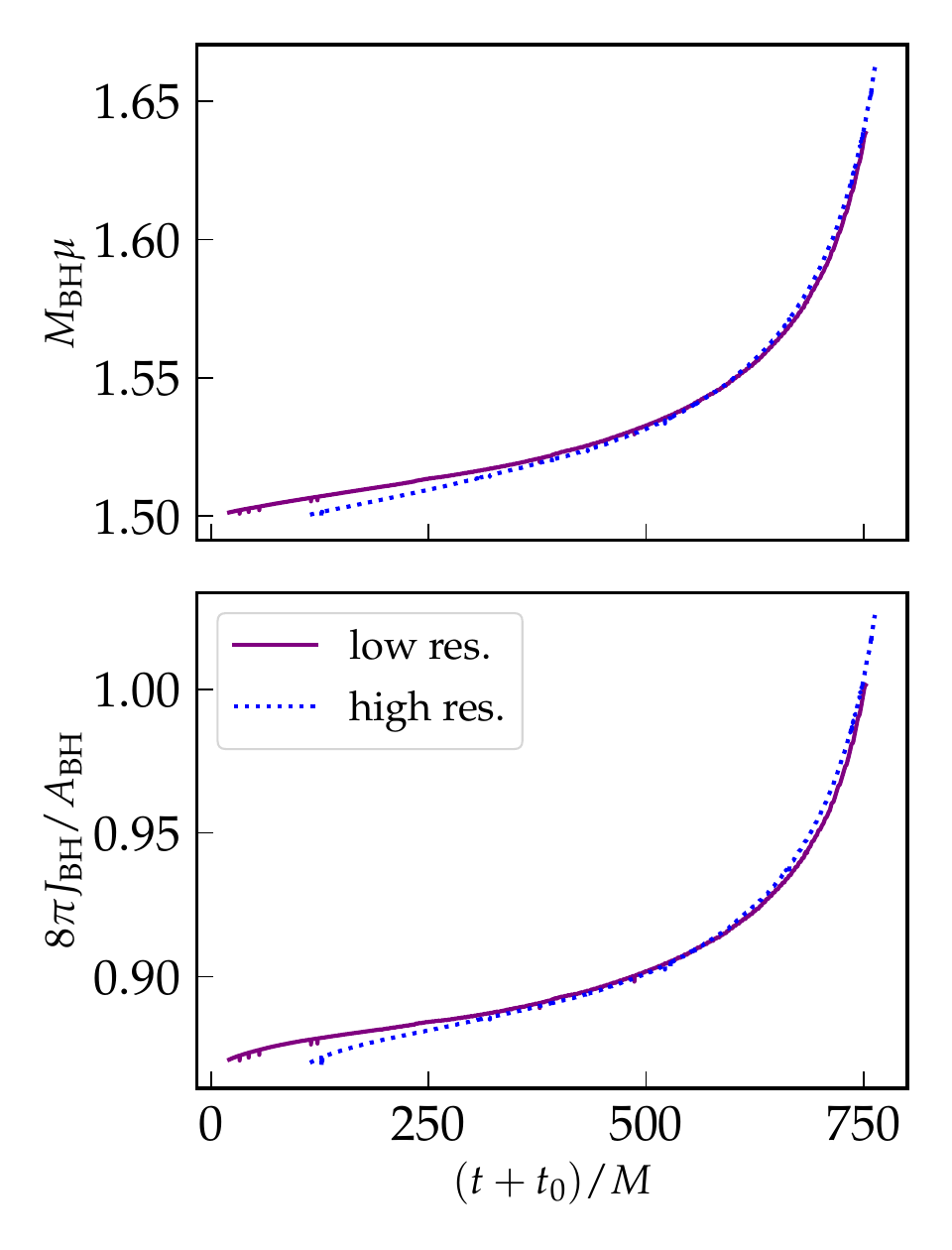}
\caption{We compare the evolution of an initial perturbation with $m=2$ and $\alpha=1.5$ on a higher resolution grid 
to a lower resolution grid with grid spacing $\approx1.3\times$ larger.  
The backreaction of the initial perturbation is dependent on the resolution of the simulation. A time shift of $t_0=100M$ is introduced  in the higher resolution curve to better compare the evolution at later times.
}
\label{fig:res_study}
\end{figure}

In Fig. \ref{fig:res_study}, we show the evolution of the same initial perturbation, with $m=2$ and $\alpha=1.5$, using two different grid resolutions. The higher resolution curve is the same one shown in Fig.~\ref{fig:mass_spin_al1p5} in the main text. The grid resolution has a noticeable impact on the initial growth of the field, changing the initial backreaction and therefore the time until the horizon reaches extremality. However, the actual growth rate $\omega_I$ is comparable for both cases in the superradiant dominated phase, corresponding to the result from the linear analysis in Ref.~\cite{East:2023nsk}, as shown for the lower resolution case in Fig. \ref{fig:growthE2al1.5}. In the figure, a relative time shift is introduced to show the comparable evolution at later times.

\begin{figure}[h]
\centering
\includegraphics[width=8cm]{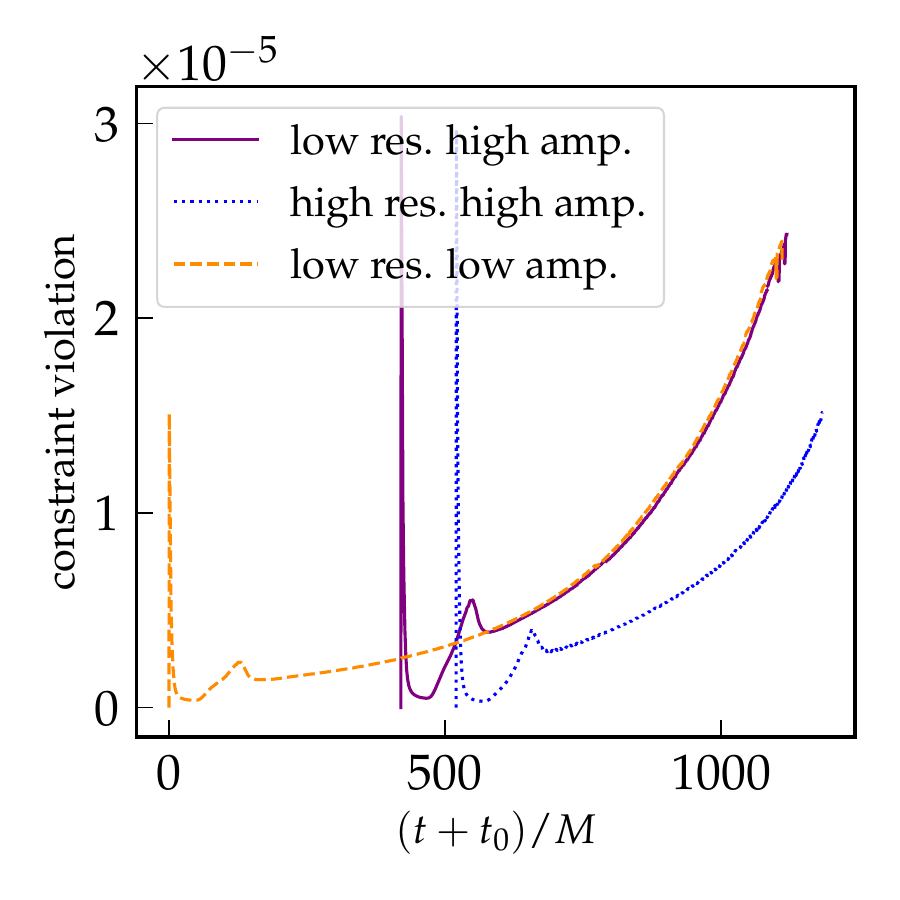}
\caption{We compare the constraint violations for all the cases plotted in Figs.~\ref{fig:diff_amps} and \ref{fig:res_study}. (The lower resolution case in Fig. \ref{fig:res_study} is the same as the higher amplitude case in Fig. \ref{fig:diff_amps}.) The start time for the higher amplitude cases is shifted, as in Figs. \ref{fig:diff_amps} and \ref{fig:res_study}, to align the background black hole evolution at later times. We have $t_0=0$, $420M$, and $520M$ for the ``low res. low amp.,'' ``low res. high amp.,'' and  ``high res. high amp.'' cases, respectively.
}
\label{fig:constraints}
\end{figure}

In Fig. \ref{fig:constraints}, we show the total constraint violation for the same two resolutions, as well as a lower resolution evolution with a lower initial amplitude. Here the constraint violation is the norm of the harmonic constraint $C^a$. From the figure, it is clear that the initial spike in the constraint violation is determined by the initial amplitude and independent of the resolution. So the initial constraint violations are dominated by the effect of using initial data which only solves the linear problem. At later times, we see that the constraint violations for the two different amplitude cases with the same resolution have the same constraint violation when aligned for the same background evolution. We also see that the constraint violation is lower for the higher resolution case at later times. This implies that the constraint violations after the first $\approx 100M$ are dominated by the truncation error. From Fig. \ref{fig:res_study}, when a time shift is included in the higher resolution run of $100M$, the mass and spin evolution line up at later time with the lower resolution simulation. Comparing the constraint violations with this time shift we find they are consistent with third order convergence.

\section{EVOLUTION IN FULL QUADRATIC GRAVITY}

In order to understand whether the second additional degree of freedom in full quadratic gravity changes the phenomenology of the nonlinear superradiant instability, we also carry out simulations evolving the Ricci scalar. Varying the action in Eq. \eqref{eq:quad_acion}, the evolution equation for the Ricci scalar is 
\begin{align}
\label{eq:Revo}
    \Box R &= m_0^2 R  ,
\end{align}
and the evolution equation for the traceless Ricci tensor is
\begin{align}
\label{eq:Rtevo}
\Box\tilde R_{ab}
    &=\mu^2\tilde R_{ab}+\frac12 g_{ab}\,\tilde R_{cd}\tilde R^{cd}
-2\tilde R^{cd}R_{acbd}
\nonumber \\
&
+\left(\frac{1}{6}+\frac{\mu^2}{3m_0^2}\right) R\,\tilde R_{ab}
+\left(\frac{1}{3}-\frac{\mu^2}{3m_0^2}\right)\,\nabla_b\nabla_a R\nonumber \\
&
+\frac{1}{12} g_{ab}(\mu^2-m_0^2)R
 \nonumber\\
    &-\kappa \left( n_a \nabla^c \tilde{R}_{bc} + n_b \nabla^c \tilde{R}_{ac} - \frac{1}{2} g_{ab} n^c \nabla^d \tilde{R}_{cd} \right).
\end{align}
In the above, the last line is an additional constraint damping term, as was described in \eqref{eq:Rabevo_EW}. We note that there are some differences between Eq. \eqref{eq:Rtevo} and Eq. (32) in \cite{Noakes:1983xd}, as well as between Eq. \eqref{eq:Rtevo} and Eq. (5) in \cite{Held:2023aap}, and that these references are also not in agreement. We find agreement between \eqref{eq:Rtevo} and Eq. (2) in \cite{Cayuso:2023dei}.

According to Eq.~\eqref{eq:Revo}, the Ricci scalar will always evolve as a massive scalar field on the spacetime. On a Kerr background, it can experience scalar superradiance. However, the associated instability timescales are orders of magnitude longer than considered in this study, and for a generic small initial perturbation, we would expect the Ricci scalar to be exponentially smaller than the massive tensor in the linear growth regime. We are interested in testing whether there is any enhanced growth of the Ricci scalar once the traceless Ricci tensor has grown large enough that it begins to backreact on the spacetime.

To that end, we consider a case where we set $m_0$ such that the scalar perturbation is in the superradiant regime, with $\alpha_0 = m_0 M_{\text{BH}} = 0.1$. We use the same values of $\mu M_{\text{BH}} = 1.5$, $\chi=0.99$, and initial spin-2 amplitude described in the main text.
We set the initial scalar perturbation to a nonrelativistic approximation for a scalar superradiant mode with $m=1$: 
\begin{align}
    &R = A\ r\ \sin(\phi)\sin(\theta)\exp\left(-\frac{1}{2}r m_0^2 M_{\text{BH}}\right),\nonumber\\
    &\partial_tR = A\ r\ m_0\cos(\phi)\sin(\theta)\exp\left(-\frac{1}{2}r m_0^2 M_{\text{BH}}\right),
\end{align}
where $A$ is the amplitude of the initial perturbation, which we set at the start of the simulation to $A = 10^{-12}/M_{\rm BH}^2$.

\begin{figure}[h]
\centering
\includegraphics[width=8cm]{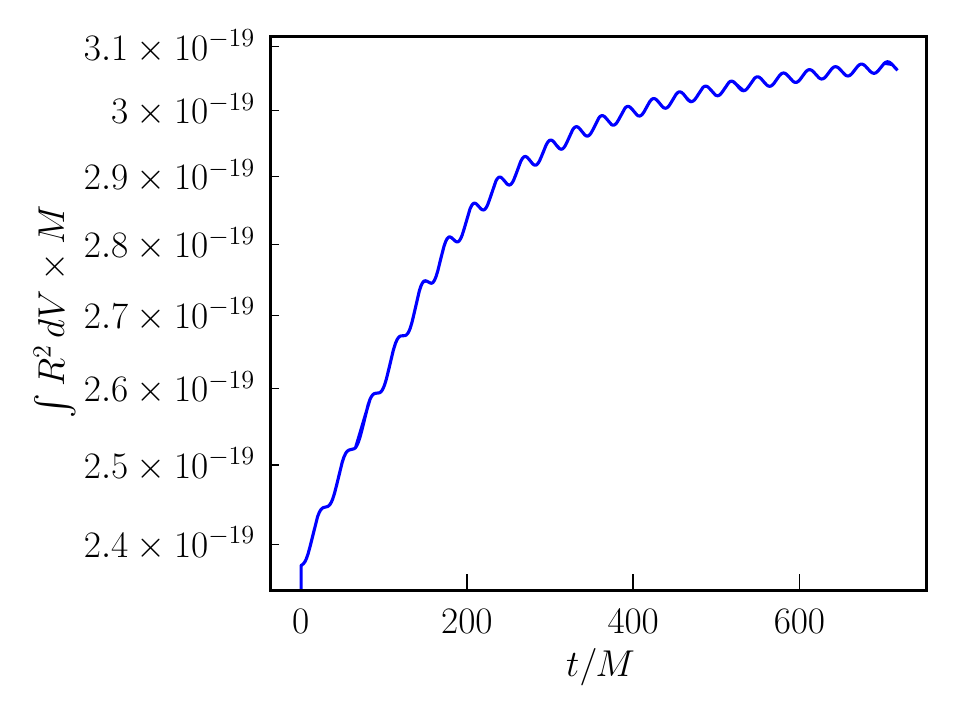}
\caption{We plot $\int dV \sqrt{-g}R^2$ as it evolves over time for a simulation that starts with $m_0 M_{\rm BH} = 0.1$ and $m_2 M_{\rm BH}=1.5$, so there is a superradiant instability in both the scalar and spin-2 degrees of freedom. In this case the black hole horizon becomes superextremal ($8\pi J_{\rm BH}/A_{\rm BH}>1$) for times $t/M > 720$. The scalar grows only slowly at late times, even in the highly nonlinear regime. 
}
\label{fig:EW_vs_quad_Rsq}
\end{figure}

Below, in Fig. \ref{fig:EW_vs_quad_Rsq}, we plot the evolution of the quantity $\int dV \sqrt{-g}R^2$. This quantifies the magnitude of $R$ throughout the evolution. As can be seen in the figure, the magnitude of the Ricci scalar does not increase significantly with time. This is expected, since the growth rate of the scalar superradiant instability is much slower than that of the spin-two superradiant instability. Inspecting the behavior of the scalar perturbation, we find that it maintains the same form as the initial perturbation, with a fixed frequency around the black hole and an $m=1$ mode structure.

The Ricci scalar stays small throughout the evolution, and has a negligible effect on the background dynamics. Testing with a much larger initial scalar amplitude ($A=10^7A_{\rm plot}$), we find the growth in $\int R^2dV$ to be exactly the same for at least the first $250M$ of evolution, when rescaled by the appropriate $1/A$ factor. This confirms that the evolution is essentially a test field evolution. The absence of any fast growth in this test field evolution indicates that, for a small initial amplitude, the scalar degree of freedom will not change the dynamics of the spin-2 superradiant instability.

\end{document}